\documentclass[english]{revtex4-1}
\usepackage[LGR,T1]{fontenc}
\usepackage[latin9]{inputenc}
\usepackage{array}
\usepackage{float}
\usepackage{textcomp}
\usepackage{multirow}
\usepackage{amsmath}
\usepackage{amssymb}
\usepackage{stmaryrd}
\usepackage{graphicx}
\usepackage{setspace}

\makeatletter

\DeclareRobustCommand{\greektext}{%
  \fontencoding{LGR}\selectfont\def\encodingdefault{LGR}}
\DeclareRobustCommand{\textgreek}[1]{\leavevmode{\greektext #1}}
\ProvideTextCommand{\~}{LGR}[1]{\char126#1}

\providecommand{\tabularnewline}{\\}

\usepackage{amsmath}

\makeatother

\usepackage{babel}
\begin{document}

\title{Future Circular Collider Based Lepton-Hadron and Photon-Hadron Colliders:
Luminosity and Physics}

\author{Y. C. Acar}
\email{ycacar@etu.edu.tr}

\affiliation{TOBB University of Economics and Technology, Ankara, Turkey }

\author{A. N. Akay}
\email{aakay@etu.edu.tr}

\affiliation{TOBB University of Economics and Technology, Ankara, Turkey }

\author{S. Beser}
\email{sbeser@etu.edu.tr}

\affiliation{TOBB University of Economics and Technology, Ankara, Turkey }

\author{A. C. Canbay}
\email{ali.can.canbay@cern.ch}

\affiliation{TOBB University of Economics and Technology, Ankara, Turkey }

\affiliation{Ankara University, Ankara, Turkey}

\author{H. Karadeniz}
\email{hande.karadeniz@giresun.edu.tr}

\affiliation{Giresun University, Giresun, Turkey}

\author{U. Kaya}
\email{ukaya@etu.edu.tr}

\affiliation{TOBB University of Economics and Technology, Ankara, Turkey }

\affiliation{Ankara University, Ankara, Turkey}

\author{B. B. Oner}
\email{b.oner@etu.edu.tr}

\affiliation{TOBB University of Economics and Technology, Ankara, Turkey }

\author{S. Sultansoy}
\email{ssultansoy@etu.edu.tr}

\affiliation{TOBB University of Economics and Technology, Ankara, Turkey }

\affiliation{ANAS Institute of Physics, Baku, Azerbaijan}
\begin{abstract}
Construction of future electron-positron colliders (or dedicated electron
linac) and muon colliders (or dedicated muon ring) tangential to Future
Circular Collider (FCC) will give opportunity to utilize highest energy
proton and nucleus beams for lepton-hadron and photon-hadron collisions.
Luminosity values of FCC based $ep$, $\mu p$, $eA$, $\mu A$, $\gamma p$
and $\gamma A$ colliders are estimated. Multi-TeV center of mass
energy $ep$ colliders based on the FCC and linear colliders (LC)
are considered in detail. Parameters of upgraded versions of the FCC
proton beam are determined to optimize luminosity of electron-proton
collisions keeping beam-beam effects in mind. Numerical calculations
are performed using a currently being developed collision point simulator.
It is shown that $L_{ep}\sim10^{32}\,cm^{-2}s^{-1}$ can be achieved
with LHeC-like upgrade of the FCC parameters.
\end{abstract}
\maketitle

\section{Introduction}

During last decades colliders have provided most of our knowledge
on fundamental constituents of matter and their interactions. Particle
colliders can be classified concerning center-of-mass energy, colliding
beams and collider types: 
\begin{itemize}
\item Collider types: ring-ring, linac-linac and linac-ring, 
\item Center-of-mass energy: energy frontiers and particle factories, 
\item Colliding beams: hadron, lepton, photon, lepton-hadron and photon-hadron
colliders. 
\end{itemize}
The ring-ring colliders are the most advanced from technology viewpoint
and are widely used around the world. As for the linac-linac colliders,
essential experience is handled due to SLC (Stanford Linear Collider
\cite{key-1} with $\sqrt{s}=0.1$ TeV) operation and ILC/CLIC (International
Linear Collider project \cite{key-2} with $\sqrt{s}=0.5-1$ TeV /
Compact Linear Collider project \cite{key-3} with $\sqrt{s}$ up
to 3 TeV) related studies. The linac-ring colliders are less familiar
(for history of linac-ring type collider proposals see \cite{key-4}).

\begin{table}[b]
\caption{Energy frontier colliders: colliding beams vs collider types.}

\begin{centering}
\begin{tabular}{|c|c|c|c|}
\hline 
Colliders  & Ring-Ring  & Linac-Linac  & Linac-Ring\tabularnewline
\hline 
\hline 
Hadron  & +  &  & \tabularnewline
\hline 
Lepton ($e^{-}$$e^{+}$)  &  & +  & \tabularnewline
\hline 
Lepton ($\mu^{-}$$\mu^{+}$)  & +  &  & \tabularnewline
\hline 
Lepton-hadron ($eh$)  &  &  & +\tabularnewline
\hline 
Lepton-hadron ($\mu h$)  & +  &  & \tabularnewline
\hline 
Photon-hadron  &  &  & +\tabularnewline
\hline 
\end{tabular}
\par\end{centering}
\centering{} 
\end{table}

In Table I we present correlations between colliding beams and collider
types for energy frontier colliders where symbol \textquotedblleft +\textquotedblright{}
implies that given type of collider provides maximal center of mass
energy for this type of colliding particles (for example; linac-ring
type colliders will give opportunity to achieve highest center of
mass energy for $ep$ collisions). Concerning the center-of-mass energy:
hadron colliders provide highest values (for this reason they are
considered as \textquotedbl{}discovery\textquotedbl{} machines), while
lepton colliders have an order smaller $E_{CM}$, and lepton-hadron
colliders provide intermediate $E_{CM}$. It should be mentioned that
differences in center-of-mass energies become fewer at partonic level.
From the BSM search point of view, lepton-hadron colliders are comparable
with hadron colliders and essentially exceeds potential of lepton
colliders for a lot of new phenomena (see \cite{key-5} for LHC (Large
Hadron Collider \cite{key-6} with $\sqrt{s}=14$ TeV at CERN), CLIC
and LEP$\varotimes$LHC (Large Electron Positron Collider \cite{key-7}
with $\sqrt{s}=0.1-0.2$ TeV at CERN) comparison and \cite{key-8}
for LHC, ILC and ILC$\varotimes$LHC comparison). 

Below we list past and future energy frontier colliders for three
time periods (Tevatron \cite{key-9} denotes $\bar{p}p$ collider
with $\sqrt{s}=1.98$ TeV at FNAL, HERA \cite{key-10} denotes $\sqrt{s}=0.3$
TeV $ep$ collider at DESY, low energy $\mu C$ denotes Muon Collider
porject \cite{key-11} with $\sqrt{s}=0.126$ TeV, LHeC denotes $\sqrt{s}=1.3$
TeV $ep$ collider project \cite{key-12}, PWFA-LC denotes Plasma
Wake-Field Accelerator-Linear Collider project \cite{key-13}, high
energy $\mu C$ denotes Muon Collider porject \cite{key-11} with
$\sqrt{s}$ up to 3 TeV): 
\begin{itemize}
\item Before the LHC (<2010): Tevatron ($\bar{p}p$), SLC/LEP ($e^{-}$$e^{+}$)
and HERA ($ep$), 
\item LHC era (2010-2030): LHC ($pp$, $AA$), ILC ($e^{-}$$e^{+}$), low
energy $\mu C$ ($\mu^{-}$$\mu^{+}$), LHeC ($ep$, $eA$) and $\mu$-LHC
($\mu p$, $\mu A$), 
\item After the LHC (>2030): FCC ($pp$, $AA$), CLIC ($e^{-}$$e^{+}$),
PWFA-LC ($e^{-}$$e^{+}$), high energy $\mu C$ ($\mu^{-}$$\mu^{+}$),
and FCC based lepton-hadron and photon-hadron colliders, namely, $e$-FCC
($ep$, $eA$) and $\mu$-FCC ($\mu p$, $\mu A$) and $\gamma$-FCC
($\gamma p$, $\gamma A$). 
\end{itemize}
Comparison of contemporary lepton and hadron colliders shows that
hadron colliders have much higher center of mass energies even at
partonic level. Therefore, formers give opportunity to search for
heavier new particles and/or probe smaller distances. This is why
they are called ``discovery'' machines. 

It is known that lepton-hadron scattering had played crucial role
in our understanding of deep inside of matter. For example, electron
scattering on atomic nuclei reveals structure of nucleons in Hofstadter
experiment \cite{key-14}. Moreover, quark parton model was originated
from lepton-hadron collisions at SLAC \cite{key-15}. Extending the
kinematic region by two orders of magnitude both in high $Q^{2}$
and small $x$, HERA (the first and still unique lepton-hadron collider)
with $\sqrt{s}=0.32$ TeV has shown its superiority compared to the
fixed target experiments and provided parton distribution functions
(PDF) for LHC and Tevatron experiments. Unfortunately, the region
of sufficiently small $x$ ($<10^{-6}$) and high $Q^{2}$ ($\geq10\,GeV^{2}$),
where saturation of parton densities should manifest itself, has not
been reached yet. Hopefully, LHeC \cite{key-12} with $\sqrt{s}=1.3$
TeV will give opportunity to investigate this region. 

Construction of linear $e^{+}e^{-}$colliders (or special linac) and
muon colliders (or special muon ring) tangential to the future circular
collider (FCC), as shown in Fig. 1, will give opportunity to achieve
highest center of mass energy in lepton-hadron and photon-hadron collisions
\cite{key-16,key-17}.

\begin{figure}[H]
\begin{centering}
\includegraphics[scale=0.5]{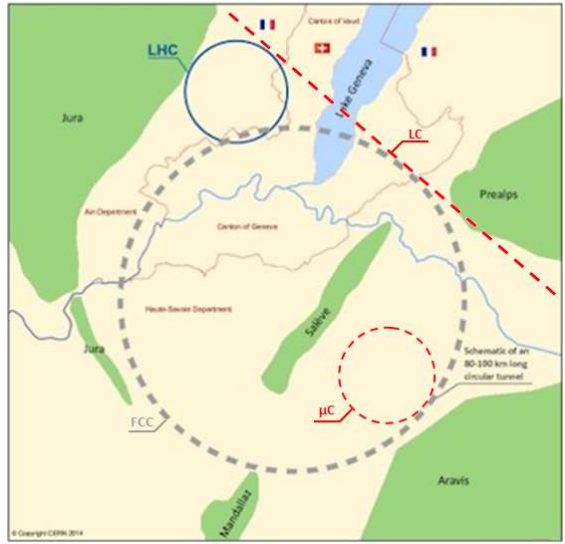} 
\par\end{centering}
\centering{}\caption{Possible configuration for FCC, linear collider (LC) and muon collider
(\textmu C). }
\end{figure}

FCC is the future 100 TeV center-of-mass energy pp collider studied
at CERN and supported by European Union within the Horizon 2020 Framework
Programme for Research and Innovation \cite{key-18}. Main parameters
of the FCC $pp$ option \cite{key-19} are presented in Table II.
The FCC also includes an electron-positron collider option in the
same tunnel (TLEP) \cite{key-20}, as well as several $ep$ collider
options \cite{key-21}. 

\begin{table}[H]
\caption{Main parameters of proton beams in FCC.}

\centering{}%
\begin{tabular}{|c|c|}
\hline 
Beam Energy (TeV)  & 50\tabularnewline
\hline 
Peak Luminosity ($10^{34}\,cm^{-2}s^{-1}$)  & 5.6\tabularnewline
\hline 
Particle per Bunch ($10^{10}$)  & 10\tabularnewline
\hline 
Norm. Transverse Emittance ($\mu m$)  & 2.2\tabularnewline
\hline 
\textgreek{b}{*} amplitude function at IP (m)  & 1.1\tabularnewline
\hline 
IP beam size ($\mu m$)  & 6.8\tabularnewline
\hline 
Bunches per Beam  & 10600\tabularnewline
\hline 
Bunch Spacing (ns)  & 25\tabularnewline
\hline 
Bunch length (mm)  & 80\tabularnewline
\hline 
Beam-beam parameter, $\xi_{pp}$ & $5.6\times10^{-3}$\tabularnewline
\hline 
\end{tabular}
\end{table}

Energy recovery linac (ERL) with $E_{e}=60\,GeV$ is chosen as the
main option for LHeC. Same ERL can also be used for FCC based $ep$
collider \cite{key-21}. Concerning $e$-ring in the FCC tunnel \cite{key-21}
energy of electrons is limited ($E_{e}<200\,GeV$) due to large synchrotron
radiation (synchrotron radiation power is proportional to the fourth
power of energy and inversely proportional to the square of the ring
radius and to the fourth power of the particle mass). Higher electron
energies can be handled only by constructing linear colliders (or
special linac) tangential to the FCC. For the first time this approach
was proposed for UNK$\varotimes$VLEPP based $ep$ colliders \cite{key-22}
(UNK denotes $pp$ collider project with $\sqrt{s}=6$ TeV at IHEP,
VLEPP denotes multi-hundred GeV $e^{+}e^{-}$ collider at BINP). Then,
construction of TESLA tangential to HERA (THERA project)  was considered
\cite{key-23}. This line was followed by consideration of the LC$\varotimes$LHC
$ep$ collider proposals (see reviews \cite{key-24,key-25,key-26}
and references therein).

In this paper, we consider main parameters of the FCC based lepton-hadron
($lp$, $lA$) and photon-hadron ($\gamma p$, $\gamma A$) colliders,
especially LC$\varotimes$FCC based ep collider schemes. In Section
II, we estimate luminosity of FCC based $ep$ colliders taking into
account beam-beam tune shift and disruption effects. In numerical
calculations, we utilize main parameters of ILC (International Linear
Collider) \cite{key-2} and PWFA-LC (Plasma Wake Field Accelerator
- Linear Collider) \cite{key-13} using a simulation program under
development for lepton-hadron colliders. Possible other options, namely,
$eA$, $\mu p/\mu A$ and $\gamma p/\gamma A$ are briefly discussed
in Section III. In Section IV, conclusions and recommendations are
presented after comparison of LC, FCC-$pp$ and LC$\varotimes$FCC
colliders' potentials for color octet electron search.

\section{LC$\varotimes$FCC Based $ep$ Colliders}

\begin{singlespace}
General expression for luminosity of FCC based $lh$ colliders is
given by ($l$ denotes electron or muon, $h$ denotes proton or nucleus):

\begin{eqnarray}
L_{lh} & = & \frac{N_{l}N_{h}}{4\pi max[\sigma_{x_{h}},\sigma_{x_{l}}]max[\sigma_{y_{h}},\sigma_{y_{l}}]}min[f_{c_{h},}\,f_{c_{l}}]\label{eq:Denklem1}
\end{eqnarray}

where $N_{l}$ and $N_{h}$ are numbers of leptons and hadrons per
bunch, respectively; $\sigma_{x_{h}}$ ($\sigma_{x_{l}}$ ) and $\sigma_{y_{h}}$
($\sigma_{y_{l}}$ ) are the horizontal and vertical hadron (lepton)
beam sizes at IP; $f_{c_{l}}$ and $f_{c_{h}}$ are LC and FCC bunch
frequencies. $f_{c}$ is expressed by $f_{c}=N_{b}f_{rep}$, where
$N_{b}$ denotes number of bunches, $f_{rep}$ means revolution frequency
for FCC and pulse frequency for LC. In order to determine collision
frequency of lh collider, minimum value should be chosen among lepton
and hadron bunch frequencies. Some of these parameters can be rearranged
in order to maximize $L_{lh}$ but one should note that there are
some main limitations that should be considered. One of these limitations
is lepton beam power, however only parameters of FCC hadron beam is
rearranged in this study and only nominal parameters of linear colliders
are considered. Therefore, there is no change of electron beam power
due to upgrades. Other limitations for linac-ring type $lh$ colliders
are due to beam-beam effects. In general, a better focusing is needed
to have high luminosity values at interaction points (IP). However,
although an intensely focused beam including charged particles with
large Lorentz factor ($\gamma$ >\textcompwordmark{}> 1) does not
have a strong influence on its internal beam particles, due to canceling
of Lorentz forces one another (space charge effects diminish with
$1/\gamma^{2}$), this situation does not hold for the encountered
beam. Deflection of particles under this electromangetic interaction
is called as disruption. When this interaction causes an angular kick
in opposite beam's particles, it is called beam-beam tune shift. While
beam-beam tune shift affects hadron (proton, ion) and muon beams,
disruption has influence on electron beams. 
\end{singlespace}

Disruption parameter for electron beam is given by: 

\begin{subequations}
\begin{eqnarray}
D_{x_{e}} & = & \frac{2\,Z_{h}N_{h}r_{e}\sigma_{z_{h}}}{\gamma_{e}\sigma_{x_{h}}(\sigma_{x_{h}}+\sigma_{y_{h}})}\label{eq:Denklem2}
\end{eqnarray}

$\,$

\begin{equation}
D_{y_{e}}=\frac{2\,Z_{h}N_{h}r_{e}\sigma_{z_{h}}}{\gamma_{e}\sigma_{y_{h}}(\sigma_{y_{h}}+\sigma_{x_{h}})}
\end{equation}

\end{subequations}

\noindent where, $r_{e}=2.82\times10^{-15}$ is classical radius for
electron, $\gamma_{e}$ is the Lorentz factor of electron beam, $\sigma_{x_{h}}$
and $\sigma_{y_{h}}$ are horizontal and vertical hadron beam sizes
at IP, respectively. $\sigma_{z_{h}}$ is bunch length of hadron beam.
$Z_{h}$ denotes atomic number for ion (for electron-proton collisions
$Z_{h}=1$). Beam-beam parameter for hadron beams is given by:

\begin{subequations}

\begin{equation}
\xi_{x_{h}}=\frac{N_{l}r_{h}\beta_{h}^{*}}{2\pi\gamma_{h}\sigma_{x_{l}}(\sigma_{x_{l}}+\sigma_{y_{l}})}\label{eq:Denklem3}
\end{equation}

$ $

\begin{equation}
\xi_{y_{h}}=\frac{N_{l}r_{h}\beta_{h}^{*}}{2\pi\gamma_{h}\sigma_{y_{l}}(\sigma_{y_{l}}+\sigma_{x_{l}})}
\end{equation}

\end{subequations}

where $r_{h}$ is radius of hadron (for proton it is classical radius,
$r_{p}=1.54\times10^{-18}$), $\beta_{h}^{*}$ is beta function of
hadron beam at interaction point (IP), $\gamma_{h}$ is the Lorentz
factor of hadron beam. $\sigma_{x_{l}}$ and $\sigma_{y_{l}}$ are
horizontal and vertical sizes of lepton beam at IP, respectively. 

Considering ILC$\varotimes$FCC and PWFA-LC$\varotimes$FCC options,
one should note that bunch spacing of electron accelerators are always
greater than FCC, while proton beam sizes are always greater than
the electron beam sizes at IP. Details and parameters of electron
beam accelerators are given in further subsections. In numerical calculations,
we use transversely matched electron and proton beams at IP. Keeping
in mind roundness of FCC proton beam, Eqs (1)-(3) turn into;

$\,$

\begin{equation}
L_{ep}=\frac{N_{e}N_{p}}{4\pi\sigma_{p}^{2}}f_{c_{e}}\label{eq:Denklem4}
\end{equation}

\begin{equation}
\xi_{p}=\frac{N_{e}r_{p}\beta_{p}^{*}}{4\pi\gamma_{p}\sigma_{p}^{2}}\label{eq:Denklem5}
\end{equation}

\begin{equation}
D_{e}=\frac{N_{p}r_{e}\sigma_{z_{p}}}{\gamma_{e}\sigma_{p}^{2}}\label{eq:Denklem6}
\end{equation}

$\,$

In order to increase luminosity of $ep$ collisions LHeC-like upgrade
of the FCC proton beam parameters have been used. Namely, number of
protons per bunch is increased 2.2 times ($2.2\times10^{11}$ instead
of $10^{11}$), $\beta$-function of proton beam at IP is arranged
to be 11 times lower (0.1 m instead of 1.1 m) which corresponds to
THERA \cite{key-23} and LHeC \cite{key-12} designs. Therefore, IP
beam size of proton beam, $\sigma_{p}$, is decreased $\sim$3.3 times
according to the relation $\sigma_{p}=\sqrt{\varepsilon_{p}^{N}\beta_{p}^{*}/\gamma_{p}}$.
Details of the parameter calculations for ILC$\varotimes$FCC and
PWFA-LC$\varotimes$FCC $ep$ colliders are given in subsections II.A
and II.B, respectively. Numerical calculations have been performed
using a new simulation software for $ep$ colliders which is currently
being developed. Details are given in subsection II.C.

\subsection{ILC$\varotimes$FCC}

Main parameters of ILC electron beam are given in Table III. One can
see from the table that bunch spacing of ILC is 554 ns which is about
22 times greater than FCC bunch spacing of 25 ns. Therefore, most
of the proton bunches turning in FCC would not participate in $ep$
collisions unless parameters of FCC (especially bunch spacing) are
rearranged. For FCC, the parameter $N_{p}$ can be increased while
number of bunches is decreased regarding the dissipation. Transverse
beam size of proton is much greater than transverse beam size of electron
for ILC$\varotimes$FCC. If beam sizes are matched, this leads $L_{ep}$
to decrease since luminosity is inversely proportional to $\sigma_{p}^{2}$
as can be seen from Eq. (\ref{eq:Denklem4}). To increase luminosity,
upgraded value of $\beta_{p}^{*}$ parameter is set to be 0.1 m and
therefore $\sigma_{p}$ to be 2.05 $\mu m$. Calculated values of
$L_{ep}$, $D_{e}$ and $\xi_{p}$ parameters for ILC$\varotimes$FCC
based $ep$ colliders with both nominal and upgraded FCC proton beam
cases are given in Table IV. In addition in Table V, disruption parameter
is fixed at the limit value of $D_{e}=25$ (for motivation see \cite{key-272727,key-282828})and
corresponding $N_{p}$ and $L_{ep}$ values are given.

\begin{table}[b]
\caption{Main parameters of electron beams in ILC \cite{key-2}.}

\centering{}%
\begin{tabular}{|c|c|c|}
\hline 
Beam Energy (GeV)  & $250$  & $500$\tabularnewline
\hline 
Peak Luminosity ($10^{34}\,cm^{-2}s^{-1}$)  & $1.47$  & $4.90$\tabularnewline
\hline 
Particle per Bunch ($10^{10}$)  & $2.00$  & $1.74$\tabularnewline
\hline 
Norm. Horiz. Emittance ($\mu m$)  & $10.0$  & $10.0$\tabularnewline
\hline 
Norm. Vert. Emittance (nm)  & $35.0$  & $30.0$\tabularnewline
\hline 
Horiz. \textgreek{b}{*} amplitude function at IP (mm)  & $11.0$  & $11.0$\tabularnewline
\hline 
Vert. \textgreek{b}{*} amplitude function at IP (mm)  & $0.48$  & $0.23$\tabularnewline
\hline 
Horiz. IP beam size (nm)  & $474$  & $335$\tabularnewline
\hline 
Vert. IP beam size (nm)  & $5.90$  & $2.70$\tabularnewline
\hline 
Bunches per Beam  & $1312$  & $2450$\tabularnewline
\hline 
Repetition Rate (Hz)  & $5.00$  & $4.00$\tabularnewline
\hline 
Beam Power at IP (MW)  & $10.5$  & $27.2$\tabularnewline
\hline 
Bunch Spacing (ns)  & $554$  & $366$\tabularnewline
\hline 
Bunch length (mm)  & $0.300$  & $0.225$\tabularnewline
\hline 
\end{tabular}
\end{table}

\begin{table}[h]
\caption{Main parameters of ILC$\varotimes$FCC based $ep$ collider.}

\noindent \centering{}%
\begin{tabular}{|c|c|c|c|c|}
\hline 
 &  & \multicolumn{3}{c|}{Nominal FCC}\tabularnewline
\hline 
$E_{e}(GeV)$  & $\sqrt{s}(TeV)$  & $L_{ep},\,cm^{-2}s^{-1}$  & $D_{e}$  & $\xi_{p}$ \tabularnewline
\hline 
250  & 7.08  & $2.26\times10^{30}$ & 1.0  & $1.09\times10^{-3}$ \tabularnewline
\hline 
500  & 10.0  & $2.94\times10^{30}$ & 0.5  & $9.40\times10^{-4}$ \tabularnewline
\hline 
$E_{e}(GeV)$  & $\sqrt{s}(TeV)$  & \multicolumn{3}{c|}{Upgraded FCC}\tabularnewline
\hline 
250  & 7.08  & $55.0\times10^{30}$ & 24  & $1.09\times10^{-3}$\tabularnewline
\hline 
500  & 10.0  & $70.0\times10^{30}$ & 12  & $9.40\times10^{-4}$\tabularnewline
\hline 
\end{tabular}
\end{table}

\begin{table}[H]
\caption{Main parameters of ILC$\varotimes$FCC based $ep$ collider corresponding
to the disruption limit $D_{e}=25$. }

\centering{}%
\begin{tabular}{|c|c|c|c|c|}
\hline 
$E_{e}(GeV)$  & $\sqrt{s}(TeV)$  & $N_{p}(10^{11})$  & $L_{ep},\,cm^{-2}s^{-1}$  & $\xi_{p}$\tabularnewline
\hline 
250  & 7.08  & 2.3  & $57\times10^{30}$ & $1.09\times10^{-3}$\tabularnewline
\hline 
500  & 10.0  & 4.6  & $149\times10^{30}$ & $9.40\times10^{-4}$\tabularnewline
\hline 
\end{tabular}
\end{table}

\subsection{PWFA-LC$\varotimes$FCC}

Beam driven plasma wake field technology made a great progress for
linear accelerators recently. This method enables an electron beam
to obtain high gradients of energy even only propagating through small
distances compared to the radio frequency resonance based accelerators
\cite{key-13}. In other words, more compact linear accelerators can
be built utilizing PWFA to obtain a specified beam energy. In Table
VI, main electron beam parameters of PWFA-LC accelerator are listed.
As in ILC$\varotimes$FCC case, transverse beam size of proton is
greater than all PWFA $e$-beam options. Same upgrade for the proton
beam is handled ($N_{p}=2.2\times10^{11}$, $\beta_{p}^{*}=0.1$ m)
and final values of luminosity, disruption and beam-beam parameters
are given in Table VII for both nominal and upgraded FCC proton beam
cases. In Table VIII, disruption parameter is fixed at the limit value
of $D_{e}=25$ and corresponding $ep$ collider parameters are given.

\begin{table}[h]
\caption{Main parameters of electron beams in PWFA-LC \cite{key-13}.}

\centering{}%
\begin{tabular}{|c|c|c|c|c|}
\hline 
Beam Energy (GeV)  & 250  & 500  & 1500  & 5000\tabularnewline
\hline 
Peak Luminosity ($10^{34}\,cm^{-2}s^{-1}$)  & 1.25  & 1.88  & 3.76  & 6.27\tabularnewline
\hline 
Particle per Bunch ($10^{10}$)  & 1  & 1  & 1  & 1\tabularnewline
\hline 
Norm. Horiz. Emittance ($10^{-5}$ m)  & $1.00$  & $1.00$  & $1.00$  & $1.00$\tabularnewline
\hline 
Norm. Vert. Emittance ($10^{-8}$ m)  & $3.50$  & $3.50$  & $3.50$  & $3.50$\tabularnewline
\hline 
Horiz. \textgreek{b}{*} function at IP ($10^{-3}$ m)  & $11$  & $11$  & $11$  & $11$\tabularnewline
\hline 
Vert. \textgreek{b}{*} function at IP ($10^{-5}$ m)  & $9.9$  & $9.9$  & $9.9$  & $9.9$\tabularnewline
\hline 
Horiz. IP beam size ($10^{-7}$ m)  & $4.74$  & $3.36$  & $1.94$  & $1.06$\tabularnewline
\hline 
Vert. IP beam size ($10^{-10}$ m)  & $26.7$  & $18.9$  & $10.9$  & $5.98$\tabularnewline
\hline 
Bunches per Beam  & 1  & 1  & 1  & 1\tabularnewline
\hline 
Repetition Rate ($10^{3}$ Hz)  & 20 & 15  & 10  & 5\tabularnewline
\hline 
Beam Power ar IP (MW)  & 8  & 12  & 24  & 40\tabularnewline
\hline 
Bunch Spacing ($10^{4}$ ns)  & $5.00$  & $6.67$  & $10.0$  & $20.0$\tabularnewline
\hline 
Bunch length ($10^{-5}$ m)  & $2.00$  & $2.00$  & $2.00$  & $2.00$\tabularnewline
\hline 
\end{tabular}
\end{table}

\begin{table}[t]
\caption{Main parameters of PWFA-LC$\varotimes$FCC based $ep$ collider.}

\centering{}%
\begin{tabular}{|c|c|c|c|c|}
\hline 
 &  & \multicolumn{3}{c|}{Nominal FCC}\tabularnewline
\hline 
$E_{e}(GeV)$  & $\sqrt{s}(TeV)$  & $L_{ep},\,cm^{-2}s^{-1}$  & $D_{e}$  & $\xi_{p}$ \tabularnewline
\hline 
250  & 7.08  & $3.44\times10^{30}$ & 1.00  & $5.47\times10^{-4}$ \tabularnewline
\hline 
500  & 10.0  & $2.58\times10^{30}$ & 0.50  & $5.47\times10^{-4}$ \tabularnewline
\hline 
1500  & 17.3  & $1.72\times10^{30}$ & 0.17  & $5.47\times10^{-4}$ \tabularnewline
\hline 
5000  & 31.6  & $0.86\times10^{30}$ & 0.05  & $5.47\times10^{-4}$ \tabularnewline
\hline 
$E_{e}(GeV)$  & $\sqrt{s}(TeV)$  & \multicolumn{3}{c|}{Upgraded FCC}\tabularnewline
\hline 
250  & 7.08  & $82.6\times10^{30}$ & 24  & $5.47\times10^{-4}$\tabularnewline
\hline 
500  & 10.0  & $61.9\times10^{30}$ & 12  & $5.47\times10^{-4}$\tabularnewline
\hline 
1500  & 17.3  & $41.3\times10^{30}$ & 4.0  & $5.47\times10^{-4}$\tabularnewline
\hline 
5000  & 31.6  & $20.8\times10^{30}$ & 1.2  & $5.47\times10^{-4}$\tabularnewline
\hline 
\end{tabular}
\end{table}

\begin{table*}[t]
\caption{Main parameters of PWFA-LC$\varotimes$FCC based $ep$ collider corresponding
to the disruption limit $D_{e}=25$. }

\centering{}%
\begin{tabular}{|c|c|c|c|c|c|}
\hline 
\multirow{1}{*}{$E_{e}(GeV)$} & \multirow{1}{*}{$\sqrt{s}(TeV)$} & \multirow{1}{*}{$N_{p}(10^{11})$} & \multirow{1}{*}{$L_{ep},\,cm^{-2}s^{-1}$} & \multirow{1}{*}{$\xi_{p}$} & \multicolumn{1}{c|}{$\tau_{IBS,x}$ (h)}\tabularnewline
\hline 
125  & 5.00  & 1.15  & $65.0\times10^{30}$ & $5.47\times10^{-4}$  & 171\tabularnewline
\hline 
250  & 7.08  & 2.30  & $86.0\times10^{30}$ & $5.47\times10^{-4}$  & 85\tabularnewline
\hline 
500  & 10.0  & 4.60  & $129\times10^{30}$ & $5.47\times10^{-4}$  & 43\tabularnewline
\hline 
1500  & 17.3  & 13.8  & $258\times10^{30}$ & $5.47\times10^{-4}$  & 14\tabularnewline
\hline 
5000  & 31.6  & 45.8  & $433\times10^{30}$ & $5.47\times10^{-4}$  & 4\tabularnewline
\hline 
\end{tabular}
\end{table*}

As one can see from the third column of the Table VIII number of protons
in bunches are huge in options corresponding to the highest energy
electron beams. Certainly, an order higher bunch population comparing
to that of FCC design value requires radical change of the injector
chain, which needs seperate study. Another critical issue is IBS growth
time. For this reason we estimate horizontal IBS growth times using
Wei formula \cite{key-27}:

$\vphantom{}$

\begin{spacing}{0}
\[
\left[\begin{array}{c}
\frac{1}{\sigma_{pf}}\frac{d\sigma_{pf}}{dt}\\
\frac{1}{\sigma_{x}}\frac{d\sigma_{x}}{dt}\\
\frac{1}{\sigma_{y}}\frac{d\sigma_{y}}{dt}
\end{array}\right]=\frac{Z^{4}Nr_{0}^{2}cL_{c}}{8\pi A\gamma^{2}\sigma_{s}\sigma_{pf}\beta\epsilon_{x}\epsilon_{y}}\times
\]

\end{spacing}

\begin{equation}
\frac{(1+a^{2}+b^{2})I(\frac{a^{2}+b^{2}}{2})-3}{1-(\frac{a^{2}+b^{2}}{2})}\left[\begin{array}{c}
(1-d^{2})\,n_{b}\\
d^{2}-(a^{2}/2)\\
-b^{2}/2
\end{array}\right]
\end{equation}

\noindent where \textit{Z} and \textit{A} are charge and atomic mass
numbers of the particle (for protons $Z=A=1$), respectively. $L_{c}$$\approx ln\left[4\beta_{rel}^{2}\bar{b}\sigma_{pf}^{2}(1-d^{2})/r_{0}(a^{2}+b^{2})\right]$
is the Coulomb logarithm factor \cite{key-303030}, $\beta_{rel}\approx1$
for ultra-relativistic particles, $a=\beta_{x}d/D_{h}\gamma$, $b=(\beta_{y}\sigma_{x}/\beta_{x}\sigma_{y})\,a$
, $d=D_{h}\sigma_{pf}/(\sigma_{x}^{2}+D_{h}^{2}\sigma_{pf}^{2})^{1/2}$
, $\sigma_{pf}$ is the fractional momentum deviation, $\sigma_{s}$
is the rms bunch length, $\sigma_{x}$ and $\sigma_{y}$ are horizontal
and vertical amplitudes, respectively. $D_{h}$ is horizontal dispersion
and its average value is equal to \cite{key-2828,key-28}:

\noindent 
\begin{equation}
\frac{l_{c}\theta_{c}}{4}(\frac{1}{sin^{2}\frac{\mu}{2}}-\frac{1}{12})
\end{equation}

\noindent where $l_{c}$ is FODO cell length and $\mu$ is the phase
advance. The bending angle per cell is taken as $\theta_{c}=2\pi/N_{c}$
where $N_{c}$ is number of FODO cells. Finally the function $I(\chi)$
is expressed as:

\begin{equation}
I(\chi)=\begin{cases}
\begin{array}{c}
\frac{1}{\sqrt{\chi(\chi-1)}}Arth\sqrt{\frac{\chi-1}{\chi}}\\
\frac{1}{\sqrt{\chi(\chi-1)}}Arctan\sqrt{\frac{1-\chi}{\chi}}
\end{array} & \begin{array}{c}
\chi\geqslant1\\
\chi<1
\end{array}\end{cases}.
\end{equation}

Obtained results for horizontal IBS growth times, $\tau_{IBS,x}=\sigma_{x}/(d\sigma_{x}/dt)$,
at $E_{p}=50$ TeV are presented in the last column of the Table VIII.
In numerical calculations we used baseline FCC FODO cell length value
$l_{c}$=203.0 m considered in \cite{key-28}. It is seen that IBS
growth times are acceptable even for $E_{e}=5000$ GeV case. 

\subsection{Collision Point Simulator for the FCC Based lepton-hadron and photon-hadron
Colliders}

There are several beam-beam simulation programs for linear $e^{+}e^{^{-}}$
and photon colliders (see for example \cite{key-29,key-30}). Unfortunately,
no similar (open-access) programs exist for $ep$ colliders. In order
to understand and analyze electron-proton beam interactions at collision
points, we start to develop a numerical program that considers beam
dynamics with aim to optimize electron and proton beam parameters
in order to obtain maximal luminosity values. At this stage luminosity,
beam-beam tune shift, disruption and beam life-time formulae (Equations
1-3, 7-9, 12-20) are included in, and the numerical results of this
paper are calculated using current software. The aim of the software
is to optimize main parameters of lepton-hadron colliders. It is obvious
that luminosity values with nominal beam parameters can be calculated
analytically. However, when beam dynamics is deeply analyzed considering
time evolution of beam structures, it becomes almost impossible to
make analytical solutions. These affects become time-dependent due
to varying beam sizes. The work on the upgraded version which will
include time dependent behaviour of beams during collision as well
as $\gamma p$ collider options is under progress. 

In addition, in order to achieve highest luminosity values at the
collision, beam parameters should be optimized. For this reason an
additional interface is being developed. It will optimize luminosity
and give required beam parameters within pre-determined parameter
interval. The current version of the program is a Java based environment
and therefore it is platform-independent. It is available to access
at http://alohep.hepforge.org and our group web page (http://yef.etu.edu.tr/ALOHEP\_eng.html). 

\section{FCC Based $\mu p$, $eA$, $\mu A$, $\gamma p$ and $\gamma A$
Colliders}

This section is devoted to brief discussion of additional options
for FCC based $lh$ and $\gamma h$ colliders.

\subsection{$\mu p$ Colliders}

Muon-proton colliders were proposed almost two decades ago. Construction
of additional proton ring in $\sqrt{s}=$ 4 TeV muon collider tunnel
was suggested in \cite{key-31} in order to handle $\mu p$ collider
with the same center-of-mass energy. However, luminosity value, namely
$L_{\mu p}=3\times10^{35}cm^{-2}s^{-1}$, was extremely over estimated,
realistic value for this option is three orders smaller \cite{key-26}.
Then, construction of additional 200 GeV energy muon ring in the Tevatron
tunnel in order to handle $\sqrt{s}=$ 0.9 TeV $\mu p$ collider with
$L_{\mu p}=10^{32}cm^{-2}s^{-1}$ was considered in \cite{key-32}.

\begin{singlespace}
In this paper we consider another design, namely, construction of
muon ring close to FCC (see Fig 1). For round beams general expression
for the luminosity given in Eq. (\ref{eq:Denklem1}) transforms to:
\begin{eqnarray}
L_{pp} & = & f_{pp}\frac{N_{p}^{2}}{4\pi\sigma_{p}^{2}}\label{eq:Denklem7}
\end{eqnarray}
 
\end{singlespace}
\begin{center}
\begin{eqnarray}
L_{\mu\mu} & = & f_{\mu\mu}\frac{N_{\mu}^{2}}{4\pi\sigma_{\mu}^{2}}\label{eq:Denklem8}
\end{eqnarray}
 
\par\end{center}

\noindent for FCC-$pp$ and $\mu C$, respectively. Concerning muon-proton
collisions one should use larger transverse beam sizes and smaller
collision frequency values. Keeping in mind that $f_{\mu\mu}$ is
an order smaller than $f_{pp}$, following correlation between $\mu p$
and $\mu\mu$ luminosities take place:
\begin{center}
\begin{eqnarray}
L_{\mu p} & = & (\frac{N_{p}}{N_{\mu}})(\frac{\sigma_{\mu}}{max[\sigma_{p},\,\sigma_{\mu}]})^{2}L_{\mu\mu}\label{eq:Denklem9}
\end{eqnarray}
\par\end{center}

\begin{table}[H]
\caption{Nominal muon collider parameters \cite{key-11}.}

\centering{}%
\begin{tabular}{|c|c|c|c|}
\hline 
$\sqrt{s}$, TeV  & 0.126  & 1.5  & 3.0 \tabularnewline
\hline 
Avg. Luminosity, $10^{34}cm^{-2}s^{-1}$  & 0.008  & 1.25  & 4.4 \tabularnewline
\hline 
Circumference, km  & 0.3  & 2.5  & 4.5 \tabularnewline
\hline 
Repetition Rate, Hz  & 15  & 15  & 12 \tabularnewline
\hline 
$\beta^{\star}$, cm  & 1.7  & 1  & 0.5 \tabularnewline
\hline 
No. muons/bunch, $10^{12}$  & 4  & 2  & 2 \tabularnewline
\hline 
No. bunches/beam  & 1  & 1  & 1 \tabularnewline
\hline 
Norm. Trans. Emmit., $\pi\:mm-rad$  & 0.2  & 0.025  & 0.025 \tabularnewline
\hline 
Bunch length, cm & 6.3 & 1 & 0.5\tabularnewline
\hline 
Beam Size at IP, $\mu m$ & 75 & 6 & 3\tabularnewline
\hline 
Beam beam parameter / IP , $\xi_{\mu\mu}$ & 0.02 & 0.09 & 0.09\tabularnewline
\hline 
\end{tabular}
\end{table}

Using nominal parameters of $\mu\mu$ colliders given in Table IX,
according to Eq. (\ref{eq:Denklem9}), parameters of the FCC based
$\mu p$ colliders are calculated and presented in Table X. Utilizing
Eq. (3) for round beams, we obtain: 

\begin{eqnarray}
\xi_{p} & = & \frac{N_{\mu}r_{p}\beta_{p}^{*}}{4\pi\gamma_{p}\sigma_{\mu}^{2}}\label{eq:Denklem10}
\end{eqnarray}

Beam beam parameter for muons is given by:

\begin{eqnarray}
\xi_{\mu} & = & \frac{N_{p}r_{\mu}\beta_{\mu}^{*}}{4\pi\gamma_{\mu}\sigma_{p}^{2}}\label{eq:Denklem11}
\end{eqnarray}

\noindent where $r_{\mu}=1.37\times10^{-17}$ m is classical muon
radius. 

\begin{table}[H]
\caption{Main parameters of the FCC based $\mu p$ colliders.}

\centering{}%
\begin{tabular}{|c|c|c|c|c|}
\hline 
Collider  & \multirow{2}{*}{$\sqrt{s}$, TeV } & $L_{\mu p},\,cm^{-2}s^{-1}$  & \multirow{2}{*}{$\xi_{p}$} & \multirow{2}{*}{$\xi_{\mu}$}\tabularnewline
Name &  & (Avg.) &  & \tabularnewline
\hline 
\hline 
$\mu63$-FCC  & 3.50  & $0.20\times10^{31}$ & $1.8\times10^{-3}$ & $5.4\times10^{-4}$\tabularnewline
\hline 
$\mu750$-FCC  & 12.2  & $49\times10^{31}$ & $1.1\times10^{-1}$ & $3.3\times10^{-3}$\tabularnewline
\hline 
$\mu1500$-FCC  & 17.3  & $43\times10^{31}$ & $1.1\times10^{-1}$ & $8.3\times10^{-4}$\tabularnewline
\hline 
\end{tabular}
\end{table}

As one can see from Table X, where nominal parameters of FCC proton
beam are used, $\xi_{p}$ for energy frontier $\mu p$ colliders is
unacceptably high and should be decreased to 0.01. According to Eq.
(\ref{eq:Denklem10}), this can be succeeded by decreasing of $\beta_{p}$
and/or increasing of $\sigma_{\mu}$. For example, decreasing $\beta_{p}^{*}$
from 1.1 m to 0.1 m (as in the upgraded option of proton beams considered
in Section II) seems to solve this problem. Luminosity values presented
in Table X assume simultaneous operation with $pp$ collider. These
values can be increased by an order using dedicated proton beam with
larger bunch population \cite{key-26}.

\subsection{$eA$ and $\mu A$ Colliders}

It is known that FCC also includes $Pb-Pb$ collider option \cite{key-18,key-28}.
Therefore, construction of LC and $\mu C$ tagential to FCC will provide
opportunity to handle $e$-Pb and $\mu$-Pb collisions. In order to
estimate luminosity of FCC based lepton-nucleus colliders we use parameters
of $Pb$-beam for $p-Pb$ option from \cite{key-28} presented in
Table XI.

\begin{table}[H]
\caption{Main parameters of $Pb$ beam in FCC $p$-$Pb$ option.}

\centering{}%
\begin{tabular}{|c|c|}
\hline 
Beam Energy (GeV)  & 4100\tabularnewline
\hline 
Peak Luminosity ($10^{30}\,cm^{-2}s^{-1}$)  & 1.24\tabularnewline
\hline 
Particle per Bunch ($10^{10}$)  & 1.15\tabularnewline
\hline 
Norm. Transverse Emittance ($\mu m$)  & 3.75\tabularnewline
\hline 
\textgreek{b}{*} amplitude function at IP (m)  & 1.1\tabularnewline
\hline 
IP beam size ($\mu m$)  & 8.8\tabularnewline
\hline 
Bunches per Beam  & 432\tabularnewline
\hline 
Bunch length (mm)  & 80\tabularnewline
\hline 
Beam-beam parameter, $\xi_{pp}$ & $3.7\times10^{-4}$\tabularnewline
\hline 
\end{tabular}
\end{table}

Luminosity, disruption and beam beam tune shift for $e$-$Pb$ are
given by:

\begin{eqnarray}
L_{ePb} & = & \frac{N_{e}N_{Pb}}{4\pi\sigma_{Pb}^{2}}f_{c_{e}}\label{eq:Denklem12}
\end{eqnarray}

\begin{eqnarray}
D_{e} & = & \frac{Z_{Pb}N_{Pb}r_{e}\sigma_{z_{Pb}}}{\gamma_{e}\sigma_{Pb}^{2}}\label{eq:Denklem13}
\end{eqnarray}

\begin{eqnarray}
\xi_{Pb} & = & \frac{N_{e}r_{Pb}\beta_{Pb}^{*}}{4\pi\gamma_{Pb}\sigma_{Pb}^{2}}\label{eq:Denklem14}
\end{eqnarray}

\noindent respectively. In Eq. (\ref{eq:Denklem14}) $\gamma_{Pb}=E_{Pb}/m_{Pb}$
and $r_{Pb}=(Z_{Pb}^{2}/A_{Pb})$r$_{p}$. Calculated luminosity values
for LC$\varotimes$FCC based e-Pb colliders are given in Table XII
(here upgraded FCC means $\beta_{Pb}^{*}=0.1$ m). One can see that
sufficiently high luminosities can be achieved with reasonable $D_{e}$
and $\xi_{Pb}$ values.

\begin{table}[H]
\caption{Main parameters of LC$\varotimes$FCC based $e$-$Pb$ colliders.}

\noindent \centering{}{}%
\begin{tabular}{|c|c|c|c|c|}
\hline 
 &  & \multicolumn{3}{c|}{Nominal FCC}\tabularnewline
\hline 
Collider  & \multirow{2}{*}{$E_{e}(GeV)$ } & \multirow{2}{*}{$L_{ep},\,cm^{-2}s^{-1}$ } & \multirow{2}{*}{$D_{e}$ } & \multirow{2}{*}{$\xi_{Pb}$ }\tabularnewline
Name &  &  &  & \tabularnewline
\hline 
\multirow{2}{*}{ILC$\varotimes$FCC} & 250  & $6.1\times10^{28}$ & 2.2 & 0.021\tabularnewline
\cline{2-5} 
 & 500  & $8.0\times10^{28}$ & 1.1 & 0.019\tabularnewline
\hline 
\multirow{4}{*}{PWFA-LC$\varotimes$FCC} & 250 & $9.4\times10^{28}$ & 2.2 & 0.011\tabularnewline
\cline{2-5} 
 & 500 & $7.0\times10^{28}$ & 1.1 & 0.011\tabularnewline
\cline{2-5} 
 & 1500 & $4.7\times10^{28}$ & 0.4 & 0.011\tabularnewline
\cline{2-5} 
 & 5000 & $2.3\times10^{28}$ & 0.1 & 0.011\tabularnewline
\hline 
 &  & \multicolumn{3}{c|}{Upgraded FCC}\tabularnewline
\hline 
Collider  & \multirow{2}{*}{$E_{e}(GeV)$ } & \multirow{2}{*}{$L_{ep},\,cm^{-2}s^{-1}$ } & \multirow{2}{*}{$D_{e}$ } & \multirow{2}{*}{$\xi_{Pb}$ }\tabularnewline
Name &  &  &  & \tabularnewline
\hline 
\multirow{2}{*}{ILC$\varotimes$FCC} & 250  & $68\times10^{28}$ & 24.5 & 0.021\tabularnewline
\cline{2-5} 
 & 500  & $88\times10^{28}$ & 12.2 & 0.019\tabularnewline
\hline 
\multirow{4}{*}{PWFA-LC$\varotimes$FCC} & 250 & $103\times10^{28}$ & 24 & 0.011\tabularnewline
\cline{2-5} 
 & 500 & $77\times10^{28}$ & 12 & 0.011\tabularnewline
\cline{2-5} 
 & 1500 & $51\times10^{28}$ & 4 & 0.011\tabularnewline
\cline{2-5} 
 & 5000 & $26\times10^{28}$ & 1.2 & 0.011\tabularnewline
\hline 
\end{tabular}
\end{table}

Luminosity and beam beam tune shifts for $\mu$-$Pb$ colliders are
given by:
\begin{center}
\begin{eqnarray}
L_{\mu Pb} & = & (\frac{N_{Pb}}{N_{\mu}})(\frac{\sigma_{\mu}}{max[\sigma_{Pb},\,\sigma_{\mu}]})^{2}L_{\mu\mu}\label{eq:Denklem15}
\end{eqnarray}
 
\par\end{center}

$\,$

\begin{eqnarray}
\xi_{\mu} & = & \frac{Z_{Pb}N_{Pb}r_{\mu}\beta_{\mu}^{*}}{4\pi\gamma_{\mu}\sigma_{Pb}^{2}}\label{eq:Denklem16}
\end{eqnarray}

\begin{eqnarray}
\xi_{Pb} & = & \frac{N_{\mu}r_{Pb}\beta_{Pb}^{*}}{4\pi\gamma_{Pb}\sigma_{\mu}^{2}}\label{eq:Denklem17}
\end{eqnarray}

\noindent Calculated luminosity values for $\mu C$$\varotimes$FCC
based $\mu$-$Pb$ colliders with nominal parameters are given in
table XIII. It is seen that nominal parameters lead to unacceptably
high $\xi_{Pb}$ values. The straightforward way to reduce $\xi_{Pb}$
is essential decreasing of $N_{\mu}$. According to Eq. (\ref{eq:Denklem15})
this leads to correspoding decreasing of luminosity as seen from the
last column of Table XIII. 

\begin{table}[H]
\centering{}\caption{Main parameters of $\mu$C$\varotimes$FCC based $\mu$-$Pb$ colliders.}
\begin{tabular}{|c|c|c|c|c|}
\hline 
 &  & \multicolumn{3}{c|}{Nominal parameters}\tabularnewline
\hline 
Collider & \multirow{2}{*}{$E_{\mu},\:TeV$ } & $L_{\mu Pb},\,cm^{-2}s^{-1}$  & \multirow{2}{*}{$\xi_{Pb}$} & \multirow{2}{*}{$\xi_{\mu}$}\tabularnewline
Name &  & (Avg.) &  & \tabularnewline
\hline 
$\mu63$-FCC  & 0.063  & $1.1\times10^{31}$ & 0.1 & $1.5\times10^{-1}$\tabularnewline
\hline 
$\mu750$-FCC  & 0.75  & $1.3\times10^{31}$ & 12 & $7.3\times10^{-3}$\tabularnewline
\hline 
$\mu1500$-FCC  & 1.5  & $1.1\times10^{31}$ & 47 & $1.8\times10^{-3}$\tabularnewline
\hline 
 &  & \multicolumn{3}{c|}{Upgraded parameters}\tabularnewline
\hline 
Collider & \multirow{2}{*}{$E_{\mu},\:TeV$ } & $L_{\mu Pb},\,cm^{-2}s^{-1}$  & \multirow{2}{*}{$\xi_{Pb}$} & \multirow{2}{*}{$N_{\mu}$}\tabularnewline
Name &  & (Avg.) &  & \tabularnewline
\hline 
$\mu63$-FCC  & 0.063  & $110\times10^{28}$ & 0.01 & $4\times10^{11}$\tabularnewline
\hline 
$\mu750$-FCC  & 0.75  & $1.1\times10^{28}$ & 0.01 & $1.67\times10^{9}$\tabularnewline
\hline 
$\mu1500$-FCC  & 1.5  & $0.23\times10^{28}$ & 0.01 & $4.26\times10^{8}$\tabularnewline
\hline 
\end{tabular}
\end{table}

\begin{figure*}[t]
\centering{}\includegraphics[scale=0.4]{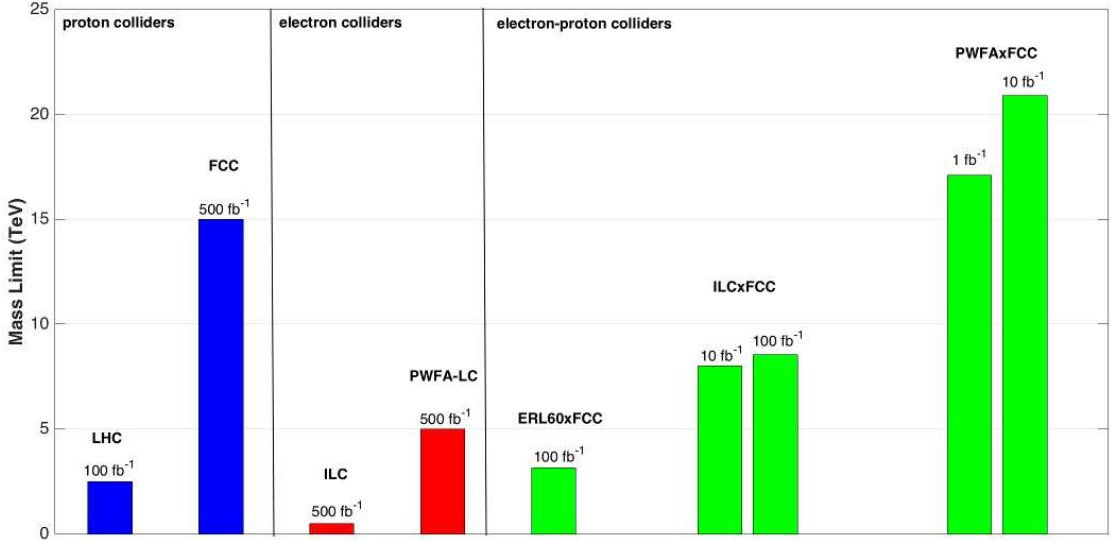}\caption{Discovery limits for color octet electron at different pp, $e^{+}e^{-}$
and $ep$ colliders.}
\end{figure*}

\subsection{$\gamma p$ and $\gamma A$ Colliders}

In 1980\textquoteright s, the idea of using high energy photon beams,
obtained by Compton backscattering of laser light off a beam of high
energy electrons, was considered for $\gamma e$ and $\gamma\gamma$
colliders (see review \cite{key-33} and references therein). Then
the same method was proposed for constructing $\gamma p$ colliders
on the base of linac-ring type $ep$ machines in \cite{key-34}. Rough
estimations of the main parameters of $\gamma p$ collisions are given
in \cite{key-35}. The dependence of these parameters on the distance
between conversion region (CR) and interaction point (IP) was analyzed
in \cite{key-30}, where some design problems were considered. 

It should be noted that $\gamma p$ colliders are unique feature of
linac-ring $ep$ colliders and could not be constructed on the base
of standard ring-ring type $ep$ machines (for arguments see \cite{key-35,key-36}).
Concerning FCC based $\gamma p$ colliders, center of mass energy
and luminosity are approximately the same as of corresponding $ep$
colliders ($\sqrt{s_{\gamma p}}\thickapprox0.9\sqrt{s_{ep}}$; $L_{\gamma p}\thickapprox L_{ep}$)
for one-pass linacs. Let us mention that energy recovery is not effective
for $\gamma p$ colliders since electron bunches are destroyed during
conversion (for details see \cite{key-36}).

Regarding the analyses performed for THERA and LHeC, $\gamma p$ colliders
have shown their superiority compared to the corresponding ep colliders
for a lot of SM and BSM phenomena (small $x_{g}$, $q^{*}$ and so
on). Similar studies should be performed for FCC based $\gamma p$
colliders. Certainly, FCC based $\gamma A$ colliders will bring out
great opportunities for QCD and nuclear physics research. For example,
$\gamma A$ option will give an opportunity to investigate quark-gluon
plasma at very high temperatures but relatively low nuclear density
(according to VMD, proposed machine will be at the same time $\rho$-nucleus
collider). 

Different aspects of the THERA based $\gamma p$ colliders have been
considered in \cite{key-37}. In \cite{key-38,key-39} Linac$\varotimes$LHC
based $\gamma p$ colliders have been considered for different linac
scenarios. Similar work on FCC based $\gamma p$ and $\gamma A$ colliders
is under progress.

\section{Conclusions}

\noindent In this study it is shown that for ILC$\varotimes$FCC and
PWFA-LC$\varotimes$FCC based $ep$ colliders, luminosity values up
to $L_{ep}\sim10^{32}\,cm^{-2}s^{-1}$ are achievable with LHeC-like
upgrade of the FCC proton beam. Even with this moderate luminosity,
BSM search potential of $ep$ colliders essentially exceeds that of
corresponding linear colliders. It may also exceed the search potential
of the FCC-$pp$ option for a lot of BSM phenomena. As a BSM process
production of color octet electron ($e_{8}$) at the FCC, LC$\varotimes$FCC
and LC have been analyzed in \cite{key-40}. Mass discovery limits
for $e_{8}$ in $\Lambda=M_{e8}$ case (where $\Lambda$ is compositeness
scale) are presented in Figure 2. If FCC will discover $e_{8}$, LC$\varotimes$FCC
will give opportunity to determine Lorentz structure of $e_{8}$-e-g
vertex using longitudinal polarization of electron beam, as well as
to probe compositeness scale up to hundreds TeV. 

In principle, ``dynamic focusing'' scheme \cite{key-41}, which
was proposed for THERA, could provide $L_{ep}\sim10^{33}\,cm^{-2}s^{-1}$
for all ep collider options considered in this study. Concerning ILC$\varotimes$FCC
based $ep$ colliders, a new scheme for energy recovery proposed for
higher-energy LHeC (see Section 7.1.5 in \cite{key-12}) may give
an opportunity to increase luminosity by an additional one or two
orders, resulting in $L_{ep}$ exceeding $10^{34}\,cm^{-2}s^{-1}$.
Unfortunately, this scheme can not be applied at PWFA-LC$\varotimes$FCC. 

Acceleration of ion beams at the FCC will give opportunity to provide
multi-TeV center of mass energy in electron-nucleus collisions. In
addition, electron beam can be converted to high energy photon beam
using Compton back-scattering of laser photons which will give opportunity
to construct LC$\varotimes$FCC based $\gamma p$ and $\gamma A$
colliders.

In conclusion, construction of ILC and PWFA-LC tangential to the FCC
will essentially enlarge the physics search potential for both SM
and BSM phenomena. Therefore, systematic study of accelerator, detector
and physics search potential, issues of LC$\varotimes$FCC based electron-hadron
and photon-hadron colliders, as well as $\mu C$$\otimes$FCC based
muon-hadron collider, are essential to plan the future of high energy
physics. Concerning the viability of different options, ILC$\varotimes$FCC
option seems to be the most realistic one for linac-ring type $ep$
machine proposals, while viability of PWFA-LC$\varotimes$FCC and
$\mu C$$\varotimes$FCC based colliders are dependent on resolution
of technical aspects of PWFA-LC and muon collider. Possible construction
of dedicated $e$-linac and/or muon ring tangential to FCC requires
separate study.

\section*{Acknowledgments}

This study is supported by TUBITAK under the grant No 114F337. A.
Akay and S. Sultansoy are grateful to organizers of the FCC Week 2016
for giving opportunity to present our results at this distinguished
conference.

\end{document}